# Reason and Method in Einstein's Relativity

Hisham Ghassib
ghassib@psut.edu.jo
The Princess Sumaya University for Technology
(PSUT), Jordan
2017

**Abstract:**

Relativity was Einstein's main research program and scientific project. It was an open-ended program that developed throughout Einstein's scientific career, giving rise to special relativity, general relativity and unified field theory. In this paper, we want to uncover the methodological logic of the Einsteinian program, which animated the whole program and its development, and as it was revealed in SR, GR, and unified field theory. We aver that the same methodological logic animated all these theories as Einstein's work progressed. Each of these theories contributed towards constructing Einstein's ambitious program. This paper is not a paper in the history of Relativity, but, rather, it utilizes our knowledge of this history to uncover the methodological logic of the relativity program and its development. This logic is latent in the historical narrative, but is not identical to it. We hope to show that the Einsteinian relativity project is still relevant today as a theoretical scheme, despite its failures and despite quantum mechanics.

# Introduction:

Many physicists and scholars have concluded that the last 25 years of Einstein's career were mis-spent and a sterile waste of effort. (Pais, 2008) They imply that Einstein reached a cul-de-sac in his scientific endeavor, and that his unified field theoretic program proved to be an utter failure and a flight of fancy. (Syrovatkin, 1979) The sterility and futility of Einstein's approach was clearly revealed by its brilliant opposite-- namely, quantum mechanics and quantum field theory. However, what this assessment seems to ignore is: first, the enormous conceptual and mathematical difficulties and absurdities encountered by both QM and QFT. Second, it ignores the unity of Einstein's thought throughout the many stages it passed through. As we shall show in this paper, this unity was animated by two principles-- the ontological priority of the concept of field, and the relativistic notion of universal law. Einstein's unified field theory was as much a product of this Einsteinian program as SR and GR. The enormous early successes of this program gave credence to its pursuance via the last, unified field theoretic, stage. A program that could lead to spectacular successes epitomized by SR and GR could surely be trusted to yield further successes in the attempt to unify the forces of nature and matter in one comprehensive unified field theory.

Elsewhere, I have shown that revolutionary development in physical theory comes as a result of a process of resolution of contradictions in the existing theory. (Ghassib, 1988; Ghassib, 1999) Both SR and GR were partial resolutions of the contradiction between the Newtonian concept of particle and the Maxwellian concept of field. (Ghassib 2012; Ghassib, 2014) However, they were partial resolutions.

That explains Einstein's dissatisfaction with both theories. The particle-field duality was as prominent in GR as it had been in classical physics. GR was a theory of space, time and gravitation, but it was not a theory of matter. It had to rely on other theories, including its archrival, quantum mechanics, to incorporate matter into its equations. To Einstein, that was unacceptable. A comprehensive, self-contained theory of everything demanded a comprehensive resolution of the particle field contradiction. Spacetime as a comprehensive field was Einstein's answer. Perhaps the proper answer would demand a new set of fundamental concepts that transcend both terms of the contradiction. Perhaps string theory would prove to be the ultimate resolution. However, Einstein chose the generalized spacetime continuum as the ultimate field that would encompass everything. Was it a wild goose chase? The trend of modern developments has proven that Einstein's dream was too ambitious to be realized in one go a la GR. New concepts are needed. However, Einstein's route was natural, logical and promising, and it was certainly a good and conceptually solid starting point for later developments. It may prove to be too premature to write it off as a wild failed chase that outlived its usefulness by 1955.

In this paper, we shall reconstruct the logic of the Einsteinian program in its various stages, showing it as an organic unity animated by the twin concepts of universal law and the spacetime continuum as a comprehensive field. This is not a study into the detailed history of development of Einstein's thought. Rather, it will take this detailed history as raw material to reconstruct its internal logic and animating principles. We will first deal with pre-relativity physics, its conceptual structure and internal contradictions. We will then show how some of these contradictions were resolved partially by SR, emphasizing the role of the animating principles of Einstein's project. We shall examine the triple contradiction between SR, acceleration and gravitation, and show how it was substantially resolved via GR, and the logic of GR is extracted via a study of its conceptual structure. Finally, we shall examine the limitations and contradictions of GR, and how they led Einstein towards his grand project of unified field theory.

This paper may be viewed as an attempt to rehabilitate Einstein's grand program.

**Pre-1905:**

The new physics that replaced the old Aristotelian physics was, right from the start, a field physics. (Ghassib, 1999; Ghassib, Aug. 2012) The old physics knew not the concept of material interaction between material entities. The distinguishing feature of the new physics is precisely the concept of interaction, and, in particular, field interaction, which was introduced by Kepler, and formalized and mathematized by Newton. (Bernard Cohen, 1983; 1985) To mathematize the description of nature, Newton had to critique the Cartesian project for natural philosophy, and introduce the concept of action at a distance, which proved to be a precursor and embryonic form of the concept of field. Kepler had already felt the same need to introduce the concept of field following the demise of Aristotelian cosmology. (Stephenson, 1987; Ghassib, 2017) However, despite this urgent need for the concept of field, it took more than a century and a half after Newton to give a mature formulation of this concept.

In fact, the concept of field in classical physics took two basic embryonic forms-- namely, action at a distance and the concept of ether. The story of the theory of relativity was, at least partly, the story of the replacement of these two embryonic concepts with the mature concept of field.

Classical mechanics was built and acquired its mature mathematical form without the need for the mature concept of field. Newton laid the foundations of classical mechanics with the concepts of material interaction and the concept of action at a distance. This classical theory of motion was mathematically completed principally by continental mathematicians, who managed to turn it into a genuine model and paradigm of theory construction for future work. The edifice of classical physics was built on that foundation. In particular, thermodynamics, optics and electromagnetism were built on the basis of classical mechanics using the concept of ether. Further developments led to the concept of field side by side with the older concepts of action at a distance and ether. However, 19th century physics did not realize the logical connection between these concepts and failed to realize that, logically speaking, the concept of field replaces the need for the two earlier concepts.

If we subscribe to the view that the concepts of ether and action at a distance are not fundamental concepts, but mere embryonic forms of the concept of field, classically, we are left with two ontological realities or substances-- namely, particles and fields. This gives rise to the following questions: How are the two substances related ontologically? Are they ontologically compatible with each other? They seem to simultaneously negate and affirm each other. How can we reconcile one with the other? Are they reducible to each other? Which of the two is ontologically more fundamental-- which of them has ontological priority? Can a reconciliation between the two be possible on the basis of classical physics? Can they be reconciled with each other without transforming appropriately one of them or both? Does the resolution of their internal and external contradictions entail their transformation and the whole framework that contains them? (Ghassib, 1988)

We aver that the whole subsequent development of theoretical physics has been an attempt to answer these questions. The latter represent the logic of development of physical theory in the last century.

It took a while for physicists to realize that fields were physical systems with many features in common with particle systems. In the second half of the 19th century, some of them made serious attempts to develop the concept of field and to reconcile it with the Newtonian concept of particle. However, what hampered their efforts was their adherence to the concept of ether and to the classical framework. In a sense, they chose the conservative route in resolving classical contradictions. (Ghassib, Aug. 2012) This applies to Lorentz, Thomson, Heaviside, Wien, Abraham, and even Poincare. (Miller, 1986) It was Einstein who followed the revolutionary route in resolving classical contradictions.

In dealing with light and electromagnetism, classical physics was forced to invent a concept that could not quite be fitted into its conceptual framework. In a sense, it was forced to invent the contradictory opposite of its basic concept of particle. Thus, classical physics unleashed within its framework new conceptual forces that would later shatter its very foundations.

The concepts of action at a distance and ether proved to be inadequate to deal with electrodynamic phenomena. In fact, they were self-contradictory concepts. Their function was material, but their essence was quite immaterial. This contradiction between function and essence plagued these concepts up to Einstein's renunciation of both of them. The concept of field proved to be the adequate alternative that successfully reconciled function and essence. To start with, it was thought to be a mere mathematical expression of force. However, it was soon realized that the opposite was the more likely relationship-- i.e., force was a quantitative manifestation of the deeper reality of field, and not the other way round. (Ghassib, 1988) The materiality of fields was progressively realized, until it reached its conclusive form in the theory of relativity. Fields were increasingly viewed as physical systems endowed with energy, linear momenta, angular momenta and even mass, on a par with particle systems. Thus, by the end of the 19th century, fields were to a large extent established as ontological physical realities on a par with corpuscular matter. The two ontological realms were qualitatively different, and yet held many properties and laws in common.

To start with, some tried to "explain" fields in mechanical terms, using elaborate mechanical models. (Kragh, 2012) It can safely be said that these attempts failed. Thus, the natural step forward was to try to explain corpuscular matter and gravitation in terms of the Maxwellian electromagnetic field. Even though that program ultimately failed, it was quite fruitful in that it influenced Einstein positively (Miller, 1977) and was a precursor to attempts at building adequate unified field theories. It was a premature attempt at unification, and it was realized by Einstein that much more work was demanded before such attempts could be possible. The problem implicit in classical physics in its mature form was that classical physics seemed to rest on two ontological substances: material particles and fields. These two substances were not monads, in the sense that they shared many fundamental properties, such as energy, linear momentum, angular momentum, mass or inertia, lagrangian, hamiltonian and others. They also constantly interacted with each other, exchanging many of their mechanical properties. Yet, they were qualitatively different in many aspects. Particles were localized, whilst fields were extended. Also, the basic concepts and laws governing particles were of a nature different from, and even opposed to, the concepts and laws governing fields. Particle systems were finite in terms of their degrees of freedom, unlike fields which were infinite. Particles and fields seemed to be simultaneously identical and different, or even contradictory. This was dimly realized by physicists at the end of the 19th century. (Kragh, 2012) Some tried to reduce fields to particles. others tried to reduce particles to the electromagnetic field. Einstein followed a third, truly revolutionary route, based on his realization that either one term, or both terms, of the fundamental contradiction would have to be transformed before unification was possible. This great insight informed all the subsequent development of Einstein's thought.

**1905-- Special Relativity** (Dingle, 1955; Feynman, 1977; Russell, 1997; Reichenbach, 2005; Gribbin, 2005)

In this section, we are interested in the conceptual, methodological and logical structure of special relativity as revealed in Einstein's 1905 paper on special relativity. (Stachel, 2005) By studying the underlying problematic, in the Althusserian sense (Ghassib, Aug. 2012), of this seminal paper, we have arrived at the conclusion that there are two underlying principles informing the whole conceptual edifice of special relativity-- namely, the principle of universality of physical laws and the principle of the ontological priority of fields.

Newton was the first to formulate correct universal laws of nature. (Bernard Cohen, 1983; Barbour, 2001) He succeeded in generalizing Galileo's and Kepler's laws and formulated a universal law of gravitation. These laws are applicable throughout nature-- everywhere in space and time. However, SR detects a flaw in this Newtonian universality. It poses two fundamental questions that address this issue: How are physical quantities and relations transformed as we move from one inertial frame of reference to another? That is, what is changeable and what is invariant as we undergo this operation? Also, how do we move mathematically from one frame to the other?

Implicit in SR is the view that a law of nature is not truly universal unless it be invariant as we move from one frame to another. Thus, SR imposes more stringent conditions on the universality of laws of nature and endows universality with a new broader meaning. However, in order to uphold this new meaning of universality, SR had to sacrifice the invariance of substantiality. (Jammer, 2006) Ultimately, the only genuine reality in nature is its universal laws. All are shadows save the forms-- a truly Platonic stance. All "substantial" physical quantities, such as length, duration, mass, electromagnetic fields and forces are transformable. This view enabled SR to dispense with the necessity of the ether. The Faraday experiment of electrical induction could be explained in terms of the transformation of the electromagnetic field, and there was no need to resort to the concept of ether. This logic led SR to posit the principle of SR as a fundamental postulate in physics. However, once it is considered universal and a necessary condition of universality, one is immediately faced with a contradiction. The Faraday and other experiments point to the idea that the laws of the EM field are invariant, obey the postulate of SR, and, thus, are truly universal laws. This necessarily leads to the peculiar conclusion that the speed of light is invariant. This conclusion bears on the answer of the second question as to how one moves mathematically from one inertial frame to another. For, it shows that the laws of EM transform differently from Newtonian mechanical laws as one moves from one frame to the other. How would one resolve this glaring contradiction?

Here, SR implicitly poses itself the following question: Which has ontological priority, the material particle or the field? SR clearly opts for the ontological priority of fields. As we shall show, this option proved to be fundamental to all Einstein's subsequent work, and it is a unifying principle of all his relativistic project.

The subsequent step in the 1905 SR paper was for Einstein to derive the SR transformation laws from the postulate of SR and the option of the ontological priority of fields. The next step was to consider Newton's laws of mechanics low-speed approximations and to reformulate them such that they would conform with the SR transformations. Thus, the SR paper implied that it was too early to reduce particles to the EM field, as Lorenz, Abraham and others had attempted to do, even though Einstein shared the view that fields had ontological priority over material particles. Einstein realized that many steps would have to be followed before such unification was possible.

A related result was Einstein's discovery of his energy equation in 1905. (Einstein, 1989) With this result, Einstein established that there are no absolute ontological walls separating material particles from fields. The result consolidated the materiality of fields and pointed to the ontological unity of the two "substances". Logically, that result launched the unified field theoretic quest, the quest for the ultimate substance of the universe.

**1907-- The Principle of Equivalence** (Ohanian, 1977; Norton, 1985; Miller, 1986; Einstein, 1989; Einstein, 2005; Pais, 2008)

Following the publication of SR, Einstein was confronted with the challenge of extending the principle of universality and the principle of the ontological priority of fields to be able to incorporate gravity and accelerated frames of reference in his relativistic scheme. Following Mach and his rejection of absolute space (Renn, 2007), Einstein was confronted with the challenge of extending the principle of SR to a principle of GR. He wanted to explore the possibility of formulating the laws of nature in the most universal possible way-- i. e., such that they are invariant relative to all frames of reference, both inertial and non-inertial.

On the other hand, he was also confronted with the challenge of formulating the theory of gravity as a field theory that would conform with the principle of GR. Others, such as Mie, Abraham and Nordstrom, were hankering after a field theory that would conform to the principle of SR. (Norton, 1992; Smeenk and Martin, 2007)

In view of Mach, the new gravitational theory would have to incorporate inertial forces within it. According to Mach, inertial forces had nothing to do with absolute space, but were the result of physical interactions between material bodies. This implied that gravitation and inertia were of the same class and transformable to each other, at least to Einstein. Thus, the pressing task following the formulation of SR was how to prove this implication, particularly the transformability of gravitation and inertia.

The key to solving this problem was Einstein's analysis of Faraday's electric induction experiment in his 1905 SR paper (Miller, 1977; Einstein, 2005). The idea of the transformability of electrostatic and magnetostatic fields into each other led Einstein to dispense with the concept of ether and to extending the principle of SR to electromagnetic phenomena. How could this analysis be extended to gravitation and inertia? The answer to this question was the principle of equivalence. The equality between the inertial mass and the gravitational mass made it possible to transform local uniform gravitational fields into inertial fields and vice versa. This was a necessary step towards formulating the principle of GR. Thus, the principle of equivalence showed that gravitation and inertia were different facets of the same field, just as electricity and magnetism were different facets of the same electromagnetic field.

This analysis shows clearly that Einstein was not interested in reconciling SR with gravitation by constructing a Lorentz invariant field theory of gravitation. Rather, he was after transcending SR towards GR, constructing in the process a field theory of gravitation and inertia that would conform with the principle of GR. The principle of equivalence was a continuation of SR and a bridge towards GR. Einstein was clearly guided by the principle of universality, the idea of field transformability, the principle of the ontological priority of fields and Mach's principle. Einstein translated Mach's principle into field terms, highlighting thus the problem leading to the principle of equivalence. Needless to say that Einstein struggled with the meaning of the equivalence principle, not only prior to the final formulation of GR in 1915, but even beyond 1915, and up to 1921 (Lehner). However, the significance of this principle is clear in view of the objective logic of connection between SR and GR.

**General Relativity** (Eddington, 1965; Smart, 1968; Einstein, 1973; Landau and Lifshitz, 1975; Vladimirov, 1987; Syrovatkin, 1979; Ghassib, 1999; Einstein, 2005; Thibault, 2005; 2006; Eisenstaedt, 2006; Pais, 2008; Dirac, 2008; Staley, 2008)

The principle of equivalence clearly pointed towards a general field theory of gravitation and inertia, that would obey the principle of GR. That was the path chosen by Einstein since 1907. Needless to say there were other paths at that time, that attempted to construct Lorentz-invariant field theories of gravitation, some of which attempted to do that in the context of unifying gravitation and electromagnetism (Norton, 1992). Einstein flirted with some of these attempts, but eventually opted completely for the path of GR. In this regard, Einstein wrote in 1933:

"I came a step nearer to the solution of the problem [of extending the principle of relativity] when I attempted to deal with law of gravity within the framework of the special theory of relativity. Like most writers at the time, I tried to frame a field-law for gravitation, since it was no longer possible, at least in any natural way, to introduce direct action at a distance owing to the abolition of the notion of absolute simultaneity. The simplest thing was, of course, to retain the Laplacian scalar potential of gravity, and to complete the equation of Poisson in an obvious way by a term differentiated with respect to time in such a way that the special theory of relativity was satisfied. The law of motion of the mass point in a gravitational field had also to be adapted to the special theory of relativity. The path was not so unmistakably marked out here, since the inert mass of a body might depend on the gravitational potential. In fact, this was to be expected on account of the principle of the inertia of energy." (Einstein, 1973; Darrigol, 2005).

At the turn of the 20th century, a number of leading physicists attempted to construct a Maxwell-type field theory of gravitation-- physicists such as Lorentz, Poincare, Abraham, Minkowski, Mie, Nordstrom and Laue (Norton, 1992; Kragh, 2015). Many of them sought a Maxwellian Lorentz covariant theory. Some of them even succeeded in incorporating the equality of inertial and gravitational masses in their theories. However, all those theories failed in their quest, first, because they limited themselves to Lorentz covariance, not to general covariance (Norton, 1986), second, because they limited themselves to a scalar potential, and, third, because they failed to account for certain observational facts, such as the bending of light beams near massive objects and the precession of the perihelion of Mercury.

Einstein's theory succeeded in transcending all these limits. Right from the start, he was guided by the principle of general relativity and general covariance (Lehner; Stachel, 1980), and envisaged the possibility of a vector or tensor gravitational potential. He also tested observationally his findings at each step of the development of his program.

As we have shown, the first step in the chain that led to GR was the principle of equivalence. The second step was Minkowski's spacetime continuum. The crucial step forward was the combination of the first two steps. Applying the principle of equivalence to the spacetime continuum would necessarily lead to the Riemannian nature of spacetime and to the idea that gravity curves spacetime. Since acceleration curves spacetime, and gravity is equivalent to acceleration, gravity is bound to curve spacetime. This marriage of the two steps led to a truly revolutionary notion in physics-- namely, that spacetime is not an eternal background of matter, but is a material field that interacts with other material fields. Spacetime is itself matter, and is, perhaps, the fundamental material substance. It also led to the revolutionary idea that spacetime as material field is gravity; no more, no less. The ten-component symmetric metric is no more and no less than the gravitational potential. After this crucial step, all that remained was to discover how this field interacted with other forms of matter, and how the properties of the spacetime field were related to the general properties of matter. (Norton, 1984; 1989; Renn, 1999; 2007; Darrigol, 2005)

Needless to say, this proved to be a formidable task. To use a metaphor employed by Einstein in 1913: "the task of constructing a field theory of gravitation was similar to finding Maxwell's equations exclusively on the basis of Coulomb's law of electrostatic forces, that is, without any empirical knowledge of non-static gravitational phenomena". (Renn, 2007)

Einstein constructed his field equations from a number of general principles that he probably considered necessary conditions of any theory worthy of the name. What guided him in that endeavor was his philosophical predispositions and experimental evidence. He constantly tested his constructions by deriving from them well-established approximations and limits and definite predictions. In particular, he utilized the idea of spacetime as a physical field very effectively. He had before him a wealth of geometric properties, worked out by Riemann, Clifford, Ricci, Levi Civita and Christoffel (Vladimirov, 1987; Syrovatkin, 1979), to choose from to construct the left-hand side of his field equations. He was also guided by the principle of GR , coupled to the principle of general covariance, Mach's principle, the conservation laws of energy and momentum, the form of the energy-momentum tensor of SR and Maxwell's EM theory, and the Newtonian limit of gravity (Ghassib, 1999). Right from the start, he tested his theory by trying to derive the Newtonian limit, the precession of the perihelion of Mercury and the effect of massive objects on light rays. He struggled with these principles for years to arrive at a satisfactory result and reconcile seemingly conflicting principles with each other, until he reached his destination in late 1915.

Even though Einstein became more and more oriented towards mathematics as his work progressed, he insisted on remaining in constant touch with experimental reality. In constructing GR, his gaze was constantly directed towards Newtonian gravity, and, in particular, towards the Poisson equation for the Newtonian gravitational field. In fact, he was guided by this experimentally supported equation. His aim was to construct the relativistic field equation for gravitation on the model of the Poisson equation. He wanted to generalize this equation in accordance with Riemannian geometry and the principle of general covariance. The right-hand side of the equation was quite easily generalized-- matter density was replaced by the stress-energy tensor. It was the left-hand term that posed a real challenge.

In the Entwurf paper of 1913 (Renn, 1999), Einstein and Grossmann chose the Ricci tensor for the spacetime gravitational curvature to ensure the compliance of the field equation with the principle of general covariance. However, they had to discard this field equation because it failed in yielding the correct Newtonian limit. This goes to show that the two main principles that constrained the Einsteinian construction were the principle of universality, epitomized by the principle of general covariance, the principle of GR, and the Newtonian limit. Einstein found it hard to reconcile the principle of general covariance in particular with the Newtonian limit. He made quite a number of mistakes before he could genuinely resolve this dilemma. The ingredients of Einstein's theoretical constructions were geometric objects, general philosophically and experimentally substantiated principles and past experimentally verified constraints and approximations.

**Unified Field Theory** (Einstein, 1935; 1950; Eddington, 1965; Syrovatkin, 1979; Einstein, 1981; 2005; Weinberg, 1992; Sauer, 2007; Pais, 2008; Staley, 2008; Kragh, 2015)

In this paper, we aver that Einstein's unified field theory program was the pinnacle of his relativity project, even though it is generally believed that it was a failure. We deem it as the natural conceptual consequence of his whole relativistic project. In view of our interpretation of the animating principles of Einstein's relativity-- namely, the principle of ultimate universality and the principle of the ontological priority of fields-- coupled to his discovery that spacetime is a field, and the gravitational field to boot, it was only natural for Einstein to pursue his relativistic project in the direction of constructing a unified field theory.

The Einstein field equation of GR was clearly an incomplete equation. Its left- hand side was purely geometrical and epitomized the two animating principles of the relativity project. However, the right-hand side is general relativistic only in form, but belongs in content to other theories, such as classical particle mechanics, quantum mechanics, nuclear physics and quantum field theory. The development of relativistic cosmology since 1917 bears witness to this fact, as its development was principally spurred by the development of the theory of matter. Thus, the challenge facing GR was how to incorporate organically the right-hand side of the GR field equation into the main scheme of the relativistic project. In particular, the challenge was how to account for particles in terms of fields (Einstein, 1935), and how to unify gravitation, electromagnetism and, perhaps, other fields within the context of a generalized spacetime as field. And, this was precisely what Einstein set out to do in the last thirty years of his life.

Even though Einstein mentioned the expression, unified field theory, for the first time in a paper he published in 1925, he actually dealt with the concept and subject before that date. In total, he published more than forty papers dealing with the subject. (Sauer, 2007) These papers constituted about a quarter of his original research papers, and about a half after 1920. This goes to show that the unified field theory scheme contained the very spirit of his relativistic project.

Prior to the relativity era, the central concern of theoreticians was to unify particles and fields on the basis of the Maxwell EM field. This was the route followed by Lorentz, Wien, Abraham, Mie and others. (Norton, 1992; Smeek and Martin, 2007; Sauer, 2007; Kragh, 2015) Following the construction of GR, it was only natural to consider the gravitational field equations to be the basis for unification. This was the route followed by Hilbert, Weyl, Kaluza, Einstein, Klein, Eddington and others. (Kragh, 2015) Generalized spacetime as generalized field was to be considered as the model and basis of unification.

We have to emphasize here that the Einsteinian unified field theory was not a single theory, but a theoretical program for constructing unified field theories. Einstein tapped the enormous mathematical resources of differential geometry, including affine geometry, spinor calculus and multi-dimensional spaces, in his search for a satisfactory theory that would embrace gravitation, electromagnetism, matter and quantum fields. (Sauer, 2007) Einstein experimented with many mathematical ideas to implement his program. However, the task proved to be too intractable to be satisfactorily accomplished at that time, even by a universal genius such as Einstein. It is our contention, however, that Einstein's failure to accomplish the task does not mean the final failure and demise of his unified field theory program. It is indeed a promising project that should be pursued, particularly in view of later developments and difficulties in quantum field theory, quantum gravity and string theory.

## Conclusion:

In this paper, we have delineated the unifying conceptual and methodological logic of development of Einstein's relativistic project, and shown how this project emanated from the heart of the contradictions of classical physics. We have emphasized that the Einsteinian unified field theory was a natural extension of SR and GR-- nay, that it represented the pinnacle of Einstein's relativistic project. We also emphasized that Einstein's failure to accommodate matter and quantum effects within his unified field theory framework does not mean the final failure and demise of his program, which could be promising in solving the many difficulties encountered by modern theories of gravitation and matter.